**Navigating Through Turbulence: Charting Early Careers in Weather and Climate Science**

Gan Zhang[1], Zhuo Wang[1], Kevin A. Reed[2], Lucas M. Harris[3], and Stephen W. Nesbitt[1]

[1] Department of Climate, Meteorology & Atmospheric Sciences, University of Illinois Urbana-Champaign, Urbana, Illinois

[2] School of Marine and Atmospheric Sciences, Stony Brook University, Stony Brook, New York

[3] Geophysical Fluid Dynamics Laboratory, National Oceanic and Atmospheric Administration, Princeton, NJ, USA

*Corresponding Author: Gan Zhang (gzhang13@illinois.edu)

## Abstract

The field of weather and climate science is at a pivotal moment, defined by simultaneous forces of institutional disruption and unprecedented technological advancements. While a shifting research and employment landscape has created career uncertainty, prompting many scientists to consider or pursue opportunities in the private sector, it has simultaneously spurred an expansion of the ecosystem through the emergence of new computational tools and the growing role of industry innovators and stakeholders. This perspective paper argues that this new, expanded ecosystem presents extraordinary opportunities for students and early-career professionals. We outline the emerging scientific frontiers powered by high-resolution simulations and artificial intelligence, suggest a practical path for navigating a more fluid career landscape, and propose how education and training must evolve. We argue that these changes expand rather than diminish the reader's capacity to do what drew most of us to this field: helping people prepare for what the atmosphere is about to do.

## Capsule

How students and early-career professionals can navigate and help shape a weather and climate enterprise being remade by AI, modeling innovations, and a more fluid job market.

## 1. Introduction: A New Era of Convergence

History often rhymes. After the Great Depression and World War II, many scientists were eager to make up for lost time and push the boundary of human knowledge. Among them were John von Neumann, Jule Charney, and their fellow computational and physical scientists in Princeton, New Jersey. Their collaborative work with the first general-purpose electronic computer launched the era of numerical weather prediction and climate modeling (Charney et al. 1950; Phillips 1956). This innovation created numerous scientific and other related career opportunities across the public and private sectors, in the United States and around the world. Today, we are living through a similar convergence of socio-technical forces. Again, computational innovation and physical science are converging, and this convergence is forging a new scientific epoch.

This transformation is set against a challenging backdrop of shifting priorities across government, academia, and industry sectors and institutional instability in the US (Voosen 2025; Mervis 2025). This has led talented scientists to transition from public service and academia to the private sector, or even consider leaving for other countries (Witze 2025; Aksenfeld and Ahart 2025). While such a migration is often framed as a "brain drain" for the public sector and the academia, we posit that the concurrent emergence of new tools and new industry stakeholders is a powerful catalyst for an expansion of the ecosystem of our field. This essay is written for students and early-career professionals (ECPs)—a group we define broadly, from undergraduates choosing a major through graduate students and postdoctoral researchers—and for the faculty, mentors, and managers who train and hire that group. While specifics will differ by career stage, the strategic posture, we argue, does not.

Meanwhile, it is worth being explicit about what has not changed. Since the birth of modern weather and climate science, the core identity of our field has rested on the noble mission of understanding the atmosphere well enough to predict it and protect people from adverse impacts. What has changed is the reach of that mission. Our forecasts have always been used by farmers, mariners, aviators, and emergency managers, and we have always relied on the technology of the day to deliver them. What is new is the breadth of audience and the immediacy of that use. Now that our science feeds directly into how a utility schedules operation, how an insurer prices coverage, and where a city places a cooling center, the distance between a scientific result and a protected person has shrunk. The tools have improved, but what has grown the most is leverage:

an ECP today can put useful information in front of the people who need it at a speed and scale that was not available a generation ago. The same mission and carried further and faster—that is the deeper reason we are optimistic.

Limited by the career experience of the authors, the writing is inevitably biased towards careers adjacent to physical science and modeling. The examples that we reach reflect that vantage point. Although readers whose work centers on instrumentation, operational forecasting, media and public communication, or the social science of risk may find our illustrations thinner than their subject deserves, we expect much of what follows applies across the field.

**2. The Evolving Scientific Frontier**

The fundamental development of our science is being reshaped by three emerging technological leaps:

- **High-Resolution Global Modeling:** For decades, long-range global modeling and shorter-range regional models operated in distinct silos, for both weather and climate. The advent of global storm-resolving models at km-scales is poised to break down these barriers. We can now simulate decades of global weather, explicitly representing high-impact extremes, allowing us to study their granular physical features and aggregated statistics at the same time (Satoh et al. 2019; Cheng et al. 2022; Takasuka et al. 2024; Segura et al. 2025).
- **Artificial Intelligence (AI):** Parallel to advances in physical modeling, AI is augmenting our capabilities with staggering speed and scale. For example, trained on vast observational and simulation datasets, AI models can emulate complex processes and show promise in creating seamless prediction tools that bridge weather and climate timescales (e.g., Lam et al. 2023; Kochkov et al. 2024; Bonev et al. 2025; Bodnar et al. 2025; Watt-Meyer et al. 2025).
- **Observation Advancements:** The observing system is advancing alongside the models with new instruments (e.g., phased array radar, Doppler Cloud Radar in Space) and platforms (e.g., uncrewed drone fleets, smallsat constellations) (Braun et al. 2018; Zhang et al. 2023; Wehr et al. 2023). Equipped with new technology and commercial capital, emerging participants prototyped novel operation modes such as balloon constellations that cruise around the globe for weeks and provide in-situ observations across distant regions (Zheng et al. 2025).

These technological advancements are forging new scientific frontiers. For ECPs, this means that expertise in short-term, high-impact weather is now directly applicable to long-term climate research, and vice versa. Meanwhile, observation activity is expanding beyond hands-on field operation to coordinated global operation systems. These changes create more scientifically integrated career paths. Furthermore, the science is moving beyond pure weather/climate prediction towards integrated risk assessment. With large, high-resolution climate ensembles, we can now better quantify the shifting likelihood of compound hazards—such as concurrent heatwaves and droughts—amidst natural variability and anthropogenic change. This requires a transition from a deterministic to a probabilistic mindset, especially for new practitioners and a broader set of stakeholders who rarely had to reason about a distribution rather than a number. While probabilistic thinking is not new to our field, the reach and need are undoubtedly new. Lastly, using new AI tools for analytics and communication, ECPs may connect physical science with societal risk assessment for industries like insurance, finance, and civil engineering with unprecedented efficiency and scale (e.g., Graham and Hoyer 2025; Weather Lab team 2025).

This shift raises fair philosophic and practical questions: if the frontier is increasingly integrated and application-driven, is weather and climate science becoming an engineering discipline, or fundamental physical science remains essential? Should all ECPs aspire to become interdisciplinary generalists, or is there still a place for the specialist? We think the professional world is wide enough for all these roles. However, the specialist's role is likely to shift from producing the final answer to being the person who can cross-examine the answer, judge whether it is physically defensible, and communicate effectively with stakeholders. That is not a lesser role. In a pipeline with many automated steps, it is the load-bearing role that earns trust through agency.

## 3. A New Ecosystem of Disruptive Forces

Societal needs and innovative technologies are reshaping our field’s ecosystem. We see multiple forces at work: the opening of research data, the partial democratization of modeling by AI emulators, the renewal of observing enterprise, the arrival of general-purpose generative AI, and a substantially broadened set of employers. The forces of modeling democratization and generative AI are related but distinct: the first concerns domain-specific models trained on geophysical data to predict the atmosphere and other components of the Earth system, and the second concerns general-purpose models (e.g., AI chatbot) that are changing how the work itself gets done. Some of these disruptions are durable, and some may prove temporary, while the technological shift is unlikely to reverse.

- **Open Access to Research Data**: Modern infrastructure development, such as cloud services, has expanded the societal reach of open-access data, such as EU's Copernicus Climate Change Service (C3S) and NOAA Open Data Dissemination (NODD). These critical platforms make observational data, model outputs, and other datasets more easily available. Timely, generous data sharing has helped disseminate valuable weather-climate information and enabled cutting-edge innovations, including the weather alerts pushed to cellphones and the training of AI weather forecast models.
- **Democratization of Model Development**: Trained with observational and physical simulation data, AI weather-climate models can finish large-scale simulations on several GPUs in days that once required weeks on thousands of CPUs in a national lab or were practically impossible for any research groups (e.g., Zhang et al. 2025; Zhang 2026). Exciting progress lies in the "marriage" of physical models and AI. AI augments our understanding of physics, while physical models remain essential for grounding AI tools. Both technologies fundamentally depend on high-quality observational data, requiring the continued expertise of scientists who build instruments, collect, and curate these datasets.
- **Observing Enterprise Under Renewal:** Every prediction model, physics-based or AI, needs to be confronted with observations, and every observation inherits the judgment calls of the people who built, calibrated, and quality-controlled the instruments behind it. The observing enterprise is undergoing its own transformation (Section 2), with major turnovers in instruments and personnel (NOAA 2026). The expertise that the observing enterprise demands is not interchangeable with modeling expertise. While careers built on that expertise are likely durable in the field, transitioning away from legacy technology and retraining the work force will encounter gaps and challenges.
- **General-Purpose Generative AI:** Distinct from the physical AI models, large language models are changing our work and dissemination. They write code, summarize literature, draft explanations, and increasingly execute multi-step agentic workflows. This raises the floor on productivity and lowers the cost of producing plausible-sounding, but misleading content. They make weather and climate misinformation cheap to generate at scale, which raises the premium on the credibility that trained professionals (e.g., broadcast meteorologists and emergency managers) have spent careers accumulating. Meanwhile, ECPs excessively rely on generative AI may struggle with building problem-solving ability and such credibility.

- **An Expanded Professional Ecosystem:** The set of organizations that need weather and climate expertise has broadened considerably, even though employment growth is structurally uneven. Technology companies like Google, Microsoft, and Nvidia are now major contributors, developing foundational AI models for weather. Financial institutions and insurance companies are hiring teams of scientists to build sophisticated risk models. Start-ups are emerging to build new observation systems and serve customers with vastly different operational needs. This expansion presents ECPs with more career options, creating a dynamic and growing ecosystem of opportunity. Importantly, this new ecosystem is promising for adventurers with transferable technical and soft skills.

The materialization of these opportunities is not necessarily guaranteed or smooth. Open access is a community achievement, mostly sustained by public institutions, and its future is not guaranteed. The same instability that unsettles careers at public institutions also puts at risk the crucial data infrastructure. Furthermore, the democratization of model development is uneven and has important nuances. Unlike running AI inference models, training frontier models requires significant, specialized computing resources that largely remain concentrated with a small number of laboratories and technology companies. These facts matter for an early-career reader. Even though access to compute, fast networks, and good mentorship remains uneven within and between countries (Jain et al. 2022). Even so, the forces above open new routes to high-impact work, and none of such work is confined to a handful of well-resourced institutions. For this shift to continue and empower more ECPs, open data, open models, and sustained public infrastructure are not conveniences but critical preconditions.

We also want to name the harder version of the worry, because we do not think it is answered by listing opportunities. Concerns about employment opportunities are not new for our field, even though underlying drivers differ across time (Horn et al. 1974; Hanson 1993). Knox (2023) already named the increasing automation and economic and political volatility as reasons why it was difficult to forecast the job market for bachelor's-level meteorologists. The improvements of new AI systems may ultimately enable them to absorb routine forecasting, first-pass analysis, and routine quality control, which have historically constituted entry-level work. The plausible fear is not that this field shrinks but reorganizes rapidly. For example, the field expands at the senior end and in emerging areas while the first rung on the ladder thins, so that the building gets taller and top-heavy as the ladder gets harder to climb. We take that seriously. Our suggestions rest on a simple observation: the tasks least exposed are the ones our training is actually built for: judging whether a result is physically defensible, recognizing when a model is confidently wrong, deciding what a stakeholder actually needs, and exercising judgment in situations no training set contains.

Consider a task that is now routine and would have been unusual a decade ago. An ECP is handed the output of an AI weather model and asked a deceptively simple question: can we trust this? Generating the forecast took minutes and cost almost nothing. Cross-examining whether it is trustworthy is the work. It means knowing what baseline the comparison should run against, what a reliability diagram shows that a deterministic skill score hides, and why a model can be optimal in mean-squared error and still smooth away the extremes a stakeholder cares about. It equally means knowing the reference: what the verifying observations actually measure, where a station record breaks, and what a reanalysis constrains rather than infers. Such cross-examination is equally important for weather and climate science (Wang et al. 2014; McGovern et al. 2024; Simpson et al. 2025). Producing a forecast is cheap now, and so is computing the skill scores. Knowing what they imply for the decision at hand is not. The judgment rests on the model, on the observations it verifies against, and on the domain knowledge of the person doing the evaluation. As Section 4 elaborates, those judgments have always been the substance of the discipline. They will also be the differentiator if early-career work moves up the value chain faster than curricula adjust—which is the case Section 5 makes.

## 4. Suggestions for Early-Career Professionals

To thrive in this environment, ECPs can benefit from focusing on building several core pillars. What follows is a long list, and we want to begin by defusing it. These are guideposts for a career, not a checklist for a semester or onboard training, and nobody is strong across all of them. Many relevant skills will only be gained through experience, and possibly many years of experience. Figure 1 sketches skill families and illustrative career paths through them. The useful question is not how you compare with an ideal candidate but which direction you are moving in and what your target roles actually value most. No one clears every band, and the bands themselves differ by role. Start with what you already have and then build tall and wide.

If you have completed a degree in this field, you have spent years learning to reason quantitatively about a complex physical system under uncertainty. In many of the teams you will join, that kind of reasoning is a rarer differentiator than a specific technical skill. Importantly, most positions in the field now sit at the intersection of several roles, ranging from developing new technologies to communicating life-saving information (AMS 2017). Anecdotally, departmental surveys of alumni point the same way, placing collaboration, teamwork, project management, writing and presenting, and understanding complex systems among the most used skills. Your training produced more of that than you probably credit it for. For many ECPs, the binding constraint is not a missing skill but a missing translation: no practice at explaining to a stakeholder or a hiring manager what your existing training equips you to do for them.

If you are a student working on your degree, build disciplinary depth first. Jain et al. (2022) warned that data-intensive work was displacing the foundational activities through which climate scientists build physical intuition. To tell when and where a result stops being reliable, the anchor remains physical science (e.g., the conservation principles). Computational tools can be learned on the job, but intuition for what the atmosphere can and cannot do is what stakeholders look for. Poor physical intuition is arguably harder and costlier for employers to fix after degree programs. Breadth matters most when it extends something you understand deeply, so let it follow the problems and people you want to serve rather than a generic list of technologies. The following list is not curated because we believe everyone should become a data scientist who knows some meteorology. We are highlighting emerging frontiers where aspiring ECPs, fluent in science, computation, and new tools, are increasingly rewarded for working effectively with user needs.

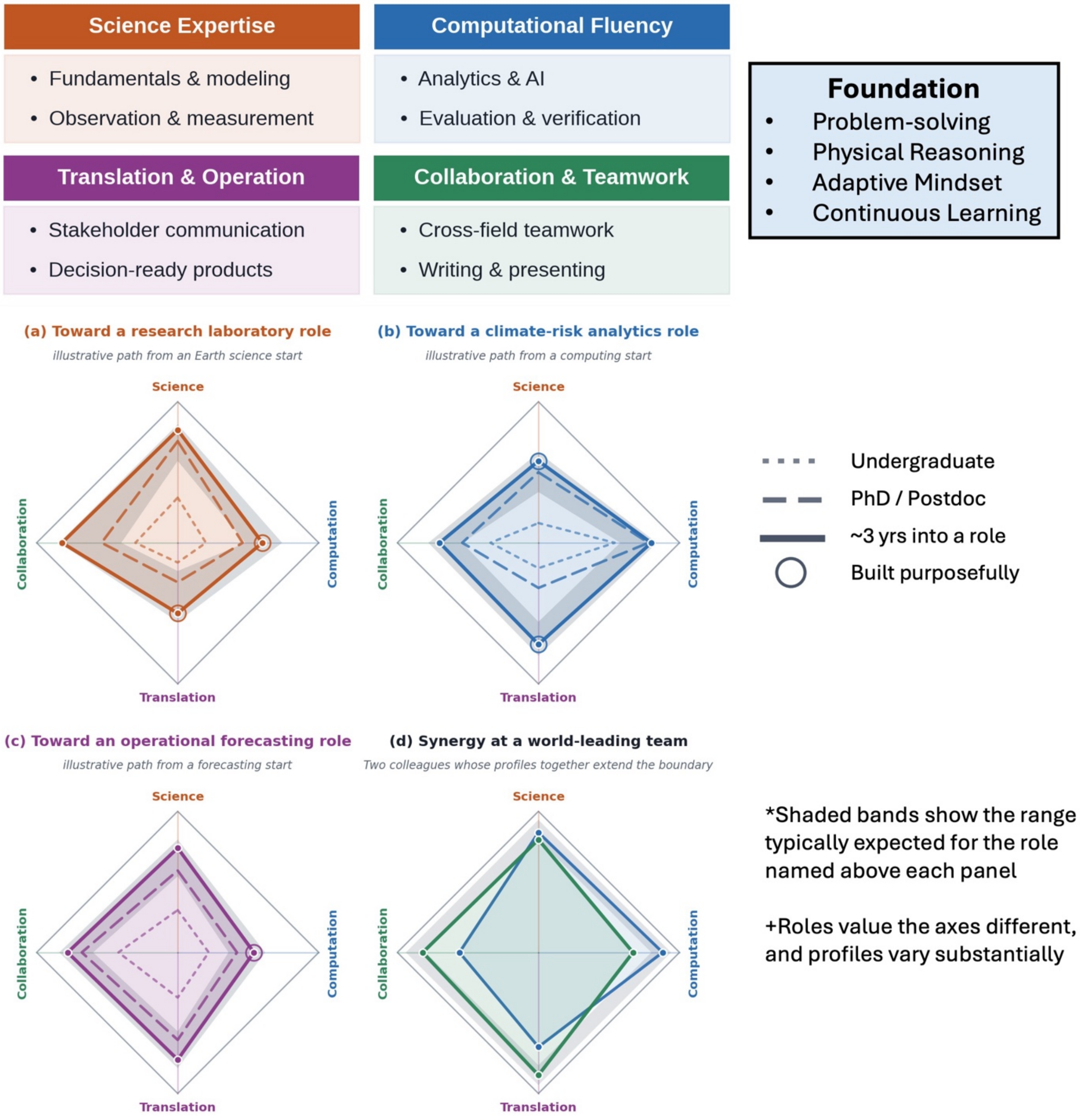

*Figure 1. Skill families discussed in section 4 (top) and illustrative career trajectories through them (bottom). Line styles denote career stage: undergraduate (dotted), PhD or postdoctoral (dashed), and approximately three years into a role (solid). Open rings mark skills built purposefully by individuals, where proficiency varies widely. Shaded bands show the range of proficiency typically expected for the role named above each panel. Panels (a)–(c) follow three entry routes: from Earth science, from computing, and from operational forecasting. Each shown against a role that route commonly leads to. Panel (d) shows an early-career scientist working alongside a colleague with exceptional computation qualifications (e.g., a computer science*

*expert). Neither profile alone spans the expertise needed for a world-leading team, but together they do. The bands differ in shape because roles value these skills differently. Each panel plots four skill families on a 0–100 proficiency scale, with the outer boundary (thin gray lines) marking 100. Values are illustrative and reflect the authors' experience rather than survey data. The pairing of starting point and role is illustrative, and each of these roles is reachable regardless of starting points.*

Importantly, you do not need to satisfy every item to apply for a job, a fellowship, or an opportunity that interests you. The evidence is that people systematically under-apply relative to their qualifications, and that this tendency is stronger among candidates from groups already underrepresented in technical fields (Coffman et al. 2024). If you are reading a position posting and counting your skill gaps, you are likely closer to qualified than you feel. Apply anyway. Early-career "failures", in the authors' personal experience and observations, often result from a lack of information and hidden learning opportunities.

*A. The Modern Scientist's Mindset*

- **Anchor Expertise in Problem-Solving, on a Foundation of Fundamentals:** The most resilient career strategy is to become genuinely good at solving problems, on top of a foundation you understand deeply. The tools will keep evolving, but the ability to formulate a problem, decompose it, and apply good judgment will remain valuable. By picking an interesting, important, and relevant problem, you will acquire the right skills as a by-product of pursuing it. This is a far more reliable route than acquiring skills first and looking for somewhere to apply them. Make sure to protect the foundation and be deliberate about AI tools. Mathematics and physics are what let you identify an artifact, and they are hard to acquire later. Generative AI can produce a working solution before you have understood the problem, which is useful for shipping but corrosive for learning.
- **Become Data and AI Literate, and Know How to Evaluate Work Output:** Take advantage of the existing open-access datasets, popular analytical tools (e.g., Python, Jupyter, and Xarray), and learn to become comfortable with working with these datasets. Understand the entire data pipeline, from the in-situ measurements collected by aircraft flying into hurricanes to the biases inherent in climate model ensembles. Be curious about the collectors and users of data, as well as their tools and work. Insights accumulated through those activities, combined with your mathematics and physics skills, underpin the judgments the scenario in Section 3 turns on, and let you evaluate work from yourself, others, and models, others. This capability matters more now that bias and misinformation can be laundered through an AI model that returns those in a convincing format.

- **Embrace Transdisciplinary Opportunities and Get Good at Working with People:** Consider cultivating a transdisciplinary foundation that blends the skills of a physicist, a modeler, and a data scientist. The goal is not necessarily to become a world-class expert in those subjects, but to develop the fluency to bridge these fields. Compared to the current generative AI, humans are distinctively good at curiosity, connection, and judgment. In practice this means the empathy to understand a stakeholder's problem, the courage to collaborate across fields with different vocabularies, and the creativity to implement new solutions not in training datasets. Decide who you want to help, find the people who serve that group, and ask what keeps them up at night. You will get a better work agenda than any list of trending techniques will give you, and often collaborators along the way.

*B. The Strategic Career Arc*

- **Embrace Career Fluidity:** Recognize that a career is not a linear path within a single sector. Rigorous training and original insights from academia or public service can provide a crucial stepping stone that launches careers in the private sector. Experience in the private sector can provide invaluable skills in product design, communication, and operational decision-making that can be brought back to academia or public service. Be open to a journey that crosses these traditional boundaries and sectors.

- **Adopt a Mindset of Continuous, Strategic Learning.** A valuable long-term asset is the ability to continuously learn by reading documentation, following tutorials, and quickly getting up to speed on new technologies. Science is moving faster than textbooks can keep up. Follow key researchers and labs on social media or network (e.g., LinkedIn), attend tech-focused meetings, read pre-prints on ArXiv, and watch conference talks online. This helps you anticipate which skills will be in demand next.

*C. The Power of the Network*

- **Build Your Community:** In a time of institutional disruption, your professional network is a great asset. Build your community proactively. This can include contributing to open-source scientific software, participating in online workshops and forums, engaging in collaborative events like hackathons, and seeking mentorship from professionals across different sectors. Professional societies including the American Meteorological Society and the American Geophysical Union maintain platforms (e.g., annual meeting events) that facilitate such interactions. These contributions and activities are a form of evidence about your passion and competence that a transcript cannot provide. Finally, actively support your peers who are navigating difficult career transitions. The scientific community is an extraordinary one, filled with people who fly *into* hurricanes while others flee. Uphold that spirit of resilience and mutual support.

We should be honest about constraints that shape career paths. For example, the doors between academia and private sectors are not equally easy to walk through in both directions. Academic hiring still credits publications more readily than industry accomplishment, so the return path from industry to academia exists but is narrower than the outbound one. Nor do our earlier advice land identically for everyone. International students make up a substantial share of graduate programs in the field, and immigration status constrains which sectors and which employers are realistically available. Relocation-heavy or low-paid early-career opportunities assume a financial cushion that many students do not have (Card et al. 2020; Stuecker et al. 2021; Card et al. 2021), while the private-sector roles we describe are geographically concentrated and not always location-flexible. None of these makes those career suggestions wrong. It does mean that individuals should weigh our suggestions accordingly and prioritize items actionable for them.

We began this section by calling it a set of guideposts rather than a checklist, and we end on the same point. No one lands at their first job holding all of these skills, and no one is expected to. Careers are built by choosing a few things that matter to you and to the people you want to serve, and by returning to the rest as circumstances evolve. If you find yourself counting what you lack, read the list again as a map of a landscape rather than preparation notes for an entrance exam.

## 5. Evolving Education for a New Generation

The challenges and opportunities of this new era demand a thoughtful evolution in how we train the next generation of scientists. While this section mainly addresses institutions and educators, we hope early-career readers will not skip it. This is also a list of things you can reasonably ask your department for, and asking is often how such changes begin.

We uphold the view that foundational physics, dynamics, and mathematics are the foundations where innovations build on. A program that adds machine learning courses by thinning the core training will risk producing graduates who can run models they cannot evaluate. With that as the constraint, we see five adaptations worth exploring. We also cite existing community work in this area, notably the AMS statements (e.g., AMS 2025) and other documents (e.g., McGovern and Allen 2021), on which several of these recommendations build.

- **Build Computation into the Required Curriculum:** While foundational physics and dynamics must remain the bedrock of our field (AMS 2017), it is essential to clear pathways for students to integrate statistics, computational science, and core computer science concepts (e.g., data structures, machine learning algorithms, and software engineering principles). For departments facing resource and other constraints, those elements can be strategically woven into the existing core courses or upper-level undergraduate electives with emphasis on career development. This integration could also be achieved through dedicated minors, interdisciplinary certificate programs, or flexible course requirements that allow for cross-listing with data science and computer science departments.
- **Embrace Project-Based Learning with Real Data and Models:** Theory must be connected to practice. Curricula should incorporate projects that integrate hands-on observation activities, use the comprehensive, publicly available datasets, and leverage user-friendly physical and AI models. Hosting campus hackathons, data analysis challenges, themed summer schools (e.g., Rossbypalooza and Graduate Climate Conference), and other summer programs (e.g., NSF LEAP and AI2ES) can provide invaluable hands-on experience to attendees and organizers. Co-organizing such activities across institutions can alleviate logistical burdens and expand platforms for ECPs to develop soft skills and professional networks (e.g., McGovern et al. 2023).
- **Foster Cross-Disciplinary and Cross-Sector Fluency:** Departments could create formal bridges to data science, statistics, and even schools of business and public policy. Economics, health, and communication are where the questions live about how risk information is received and acted on. The bridges can be strengthened via hosting joint departmental seminars and inviting speakers from the private sector to give seminars or share their first-hand experiences. Double majors outside our departments are worth actively encouraging rather than merely permitting. As with computation, an effective lever is cross-listing selected courses and making outside coursework count toward the degree. Furthermore, actively promoting and facilitating private sector internships can provide students with invaluable hands-on experience and demystify conventional career options.

- **Teach Communication and Translation Purposefully:** As our science becomes more directly tied to risk assessment and societal decision-making, the ability to communicate complex findings to non-expert stakeholders is a critical skill. Training in science communication, data visualization, and product design should be part of a modern scientist's education. For example, practice methodologies prevalent in the private sector (e.g., product thinking), translating uncertainty into risks (e.g., return on investment), and effectively communicating results to decision-makers who are *not* scientists (e.g., broadcast meteorology). Most universities already have the machinery for such training and we underuse it. Entrepreneurship centers and business engagement units routinely run training programs and are generally pleased to be invited to share their insights.

- **Encourage Community Conversations and Leadership.** The AMS statement on the bachelor's degree in atmospheric science already describes many of the highlighted skills (AMS 2017), but it was drafted before the developments in Sections 2 and 3 were visible. In the spirit of its creation process and related community conversations (e.g., Horel et al. 2013), we sketch three directions where community discussions might take. First, treat computational and statistical fluency as core rather than encouraged, with stated learning outcomes, so that departments can plan against them. Second, add a competency in evaluating model and forecast quality from increasingly diverse sources (e.g., private sector and AI emulators), together with the measurement and data-stewardship literacy that any such evaluation depends on. Third, propose recommendations to aid the graduate training, which is a cornerstone that produces our future workforce and ambassadors of this field.

We offer these as an invitation rather than a prescription, and we recognize that others in the community will weigh them differently. We also acknowledge that everything above is easier to write than to do. A faculty member teaching multiple courses every term in a program with three colleagues, facing flat or falling enrollment and a budget under review, cannot absorb this development cost by working harder. In these cases, the realistic change is not the individual department but shared infrastructure: MetEd and similar repositories, AMS professional development, and cross-institutional summer schools. Supporting and funding those bodies to develop this curriculum centrally would do more than any number of exhortations, including ours. Furthermore, the deeper institution prerequisites are structural. These changes are most likely where departments hire faculty who have worked across the public-private boundary and who practice AI-inclusive pedagogy, where such work counts toward tenure rather than against it, and where industry partnerships bring in resources and opportunities. Curriculum reform is downstream of hiring and incentives. We also recognize that this entire discussion presumes an education pipeline that is itself under strain, and that the most thoughtful curriculum cannot compensate for students who never reach it or arrive without essential preparation.

## 6. Conclusion: A Resilient Future

The field of weather and climate science is in a state of profound and challenging transition. Yet, within this disruption lies the promise of a more integrated, impactful, and dynamic future. The new ecosystem is promising and exciting, but it cannot replace the traditional system of public research and observation that forms the bedrock of our enterprise. We have encouraged you to treat sector boundaries as permeable, while maintaining that the public core is what makes the rest possible. However, permeability is healthy only if the core is sustained, and that sustaining it is partly the responsibility of those who benefit from it. For students and early-career professionals who cultivate solid foundations, embrace career fluidity, master problem-solving, and commit to supporting their community, this is not an era to fear, but one to lead.

Technological innovation and infrastructure reorganization are accelerating as we prepare this essay. The specific technologies in this essay will be superseded, probably faster than we expect. Some of what looks certain now will look like a passing phase. The durable advice is about judgment and adaptability, not about any current tool or technical skill. What will not be superseded is the reason most of us are here. The atmosphere will keep affecting people in profound ways, and the work of acting for these people will not run out. By working together, we can ensure that our field and its community thrive in the coming decades.

**Acknowledgement**

G.Z. acknowledges discussions with students, early-career professionals, experienced scientists including Drs. Gregory Hakim, Andrew Hazelton, Sarah Kapnick, Adam Sobel, Baoqiang Xiang, Janni Yuval that motivated this contribution. G.Z. is grateful for the students and faculty members of Rossbypalooza for opening their space and sharing their thoughts. G.Z. and K.R. thank Drs. Isla Simpson, Brian Medeiros, and other NSF NCAR researchers who shared their experience in mentoring students and early-career professionals. G.Z. is supported by the U.S. National Science Foundation award AGS-2327959 and RISE-2530555, as well as the faculty development fund of the University of Illinois Urbana-Champaign. K.A.R. is supported by the U.S. National Science Foundation award AGS-2327958 and the U.S. Department of Energy Office of Science through award DE-SC0025738.

This publication represents the personal views of the authors and does not necessarily reflect the policies, positions, or practices of affiliated organizations. The inclusion of the authors' affiliations and research sponsors is for identification purposes only and does not imply endorsement by the organizations.